\documentclass[prl, superscriptaddress, twocolumn, showpacs, nofootinbib]{revtex4}

\usepackage[]{graphicx}
\usepackage{amsmath}
\usepackage{amsfonts}
\usepackage{color}

\def\ket#1{| #1 \rangle}
\def\bra#1{\langle #1 |}
\def\kb#1#2{|#1\rangle\!\langle #2 |}

\def\II{1\!\mathrm{l}}

\def\Tr{{\mathrm{Tr}}}

\pacs{05.45.Mt, 03.67.Lx}
\date{\today}

\begin{document}

\title{Exponential speed-up with a single bit of quantum information:\\
Testing the quantum butterfly effect}

\author{David Poulin}
\affiliation{Institute for Quantum Computing, University of Waterloo, ON, N2L 3G1, Canada; and \\ Perimeter Institute for Theoretical Physics, 35 King
  Street N., Waterloo, ON, N2J 2W9, Canada}
\author{Robin Blume-Kohout}
\affiliation{Theoretical Division, LANL, MS-B213, Los Alamos, NM  87545, USA}
\author{Raymond Laflamme}
\affiliation{Institute for Quantum Computing, University of Waterloo, ON, N2L 3G1, Canada; and \\ Perimeter Institute for Theoretical Physics, 35 King
  Street N., Waterloo, ON, N2J 2W9, Canada}
\author{Harold Ollivier}
\affiliation{INRIA - Projet Codes, BP 105, F-78153 Le Chesnay, France}

\begin{abstract}
  We present an efficient quantum algorithm to measure the average
  fidelity decay of a quantum map under perturbation using a single
  bit of quantum information. Our algorithm scales only as the
  complexity of the map under investigation, so for those maps
  admitting an efficient gate decomposition, it provides an
  exponential speed up over known classical procedures.  Fidelity
  decay is important in the study of complex dynamical systems, where
  it is conjectured to be a signature of quantum chaos. Our result
  also illustrates the role of chaos in the process of decoherence.
\end{abstract}

\maketitle

``The flap of a butterfly's wings in Brazil can set off a tornado in
Texas''. This {\it butterfly effect} illustrates the canonical feature
of chaotic systems: they display extreme sensitivity to their initial
conditions. In a chaotic regime, phase space trajectories diverge
exponentially in time at a rate governed by the largest Lyapunov
exponent of the system. Isolated quantum systems cannot display the
butterfly effect, since unitary evolution preserves distances between
states.  Extensive research over the past two decades has been devoted
to examining other manifestations which can be used to distinguish the
regular and chaotic regimes of quantum systems. While many signatures
of quantum chaos have been proposed, their validity relies mostly on
vast accumulations of numerical evidences. Furthermore, obtaining
conclusive results using these measures requires manipulating data
whose size scales as the dimension ($N$) of the system's Hilbert space
-- that is, exponentially with the number of qubits $K$ required to
simulate the system.  In this article, we demonstrate a quantum
algorithm to evaluate one such signature ---~the average fidelity
decay~--- with a {\em single} bit of quantum information, in a time
that scales as $\mbox{poly}(K)$.

Fidelity decay was initially proposed as a signature of chaos by Peres
\cite{Peres1984}, and has since been extensively investigated
\cite{FD,JSB2001}.  The closest quantum analogue to the (purely
classical) butterfly effect, fidelity decay measures the rate at which
identical initial states diverge when subjected to slightly different
dynamics.  The discrete time evolution of a closed quantum system can
be specified by a unitary operator $U$, where $\rho(\tau_n) = U^n\rho_0
(U^{\dagger})^n$.  To examine fidelity decay, we construct a slightly
perturbed map $U_p$, where $U_p = UP$ with $P = \exp\{-i\delta V\}$
for some small $\delta$ and a hermitian matrix $V$. It is conjectured
that the overlap (or fidelity)
\begin{equation}
F_n(\psi) = \left|\bra\psi (U^n)^\dagger U_p^n \ket\psi \right|^2
\label{FD}
\end{equation}
between initially identical states $\psi$ undergoing slightly
different {\it evolutions}, $U$ and $U_p$, should decay differently
(as a function of the discrete time $n$) for regular and chaotic
dynamics: chaotic dynamics will display exponential fidelity decay,
while regular dynamics will produce polynomial fidelity decay.  Actual
results to date show behavior rather more complex than the preceding
simple conjecture.  There are various regimes governing the decay
rate; fidelity decay depends on the perturbation strength $\delta$,
and also on the degree of correlation between the eigenbasis of $U$
and the eigenbasis of the perturbation $P$~\cite{EWLC2002}.  Fidelity
decay remains a powerful diagnostic of chaotic behavior, but
calculating it is computationally hard. Furthermore, because
$F_n(\psi)$ generally shows large fluctuations over time, it is in
practice necessary to average $F_n(\psi)$ over a random set of initial
states $\psi$ to determine its decay rate, thus increasing the
numerical burden.

Several classically hard problems can be solved in polynomial (in $K$)
time on a quantum computer.  Since fully controllable and scalable
quantum computers are still quite a ways in the future, algorithms
which can be performed on a less-ambitious quantum information
processor (QIP) are of great interest.  A QIP is a quantum device
which may fail to satisfy one or more of DiVincenzo's five criteria,
but can nonetheless carry out interesting computations~\cite{BCD2002}.
Of particular interest to us is deterministic quantum computation with
a single bit (DQC1)~\cite{KL1998}, a model of quantum information
processing which is believed to be less powerful than universal
quantum computation and which is naturally implemented by a high
temperature NMR QIP~\cite{Cory2000}. In this model, universal control
over all qubits is still assumed, but state preparation and read-out
are limited. The initial state of the $(K+1)$-qubit register is
\begin{equation}
\left( \gamma \kb{0}{0} + \frac{1-\gamma}{2}\II\right) \otimes
\frac{\II}{2^{K}},
\label{eq:initial}
\end{equation} 
i.e., the first qubit (called the probe qubit for reasons which will
become clear) is in a pseudo-pure state, whereas the other $K$ qubits
are in the maximally mixed state. Furthermore, the result of the
computation is obtained as the noisy expectation value of $\sigma_z$
on the probe qubit. The variance of $\sigma_z$ is determined by {\em
i}) the polarization $\gamma$ of Eq.~\ref{eq:initial} (independent of
the size of the register) and {\em ii}) the inherent noise of the
measuring process. Hence, $\langle \sigma_z\rangle$ can be estimated
to within arbitrary $\epsilon$ with a probability of error at most $p$
by repeating the computation $O(\log(1/p)/\epsilon^2)$
times~\cite{Huber1981}. The value of $\gamma$ in high-temperature NMR
is independent of the size of the register because only a single qubit
needs to be in a pseudo-pure state. The ``inherent noise'' receives
contribution from both electronic noise and statistical fluctuations
due to the finite sample size.

While it has been known for some time that the dynamics of some
quantized chaotic systems can be {\it efficiently} (i.e., in
poly($K$) time) simulated on quantum computers~\cite{simul},
it was shown only recently \cite{ELPC2003,EWLC2002,PLMP2003} that this
ability can also be used to efficiently evaluate certain proposed
signatures of quantum chaos.  In Ref.~\cite{ELPC2003}, an efficient
quantum circuit is constructed to evaluate the coarse grained local
density of states (LDOS) ---~the average profile of the eigenstates of
$U$ over the eigenbasis of $U_p$~--- which is believed to be a valid
indicator of chaos and is formally related to fidelity decay via
Fourier transform~\cite{JSB2001}. In Ref.~\cite{EWLC2002}, an
efficient procedure to estimate the fidelity decay using the standard
model of quantum computation is presented. Finally, in
Ref.~\cite{PLMP2003}, a DQC1 circuit is presented to estimate the
form factors $t_n = \left|\Tr\{U^n\}\right|^2$ of a unitary map $U$
which, under the random matrix conjecture (see \cite{Haake2001} and
refs. therein), is a good signature of quantum chaos. The
proposed algorithm offers only a quadratic speedup, but since
entanglement is very limited in DQC1~\cite{ZHSL1998_BCJLPS1999}, this
result raises doubt about the common belief that massive entanglement
is responsible for quantum computational speed-up~\cite{EJ1998}.

Drawing upon all this previous work, we will now construct an
efficient DQC1 algorithm to evaluate the average fidelity decay
associated with any pair of unitary operators $U$ and $U_p$, provided
they can be implemented efficiently, e.g. as those of
Refs.~\cite{simul}. We begin by proving a crucial identity required to
implement the efficient algorithm.

Let $f(\psi)$ be a complex-valued function on the space of pure state
of a $N$-dimensional quantum system. We denote its average by
$\overline{f(\psi)} = \int f(\psi) d\psi$, where $d\psi$ is the
uniform measure induced by the Haar measure, such that $\int d\psi =
1$. For sake of compactness let $\bra\psi A \ket \psi = \langle A
\rangle_\psi $.

\noindent{\bf Theorem:}~Let $A,B,C,\ldots$ be $\ell$ linear operators on 
a $N$-dimensional Hilbert space. Then
\begin{equation}
\overline{ \langle A \rangle_\psi \langle B \rangle_\psi \langle C \rangle_\psi
\ldots} = \frac{\Tr\left\{(A\otimes B\otimes C \ldots) P_S^{(\ell)}\right\}}
{{N+\ell-1 \choose \ell}}
\label{theorem}
\end{equation}
where $P_S^{(\ell)}$ is the projector on the symmetric subspace of
$\ell$ systems, see Ref.~\cite{BBDEJM1997} for details on $P_S^{(\ell)}$.

\noindent{\bf Proof:}~First, note that
\begin{equation*}
\langle A \rangle_\psi \langle B \rangle_\psi \langle C \rangle_\psi \ldots
= \Tr\left\{\kb\psi\psi^{\otimes \ell} (A\otimes B\otimes C \ldots)\right\}.
\end{equation*}
Therefore, the average over the pure states $\psi$ yields,
\begin{equation*}
\Tr\left\{\overline{\kb\psi\psi^{\otimes \ell}} 
(A\otimes B\otimes C \ldots)\right\}.
\end{equation*}
Since $\overline{\kb\psi\psi^{\otimes \ell}}$ annihilates any state
which is antisymmetric under interchange of two of the $\ell$ systems,
and is by construction symmetric under such interchange, it must be
proportional to the projector $P_S^{(\ell)}$ onto the symmetric
subspace. To establish the theorem it is sufficient to find the
proportionality factor $\lambda$ between these two quantities. Letting
$A=B=C=\ldots = \II$, we get $1 = \Tr\{\overline{\kb\psi\psi^{\otimes
\ell}}\} = \lambda \Tr\{P_S^{(\ell)}\} = \lambda {N+\ell-1 \choose
\ell}$ (see Ref.~\cite{BBDEJM1997}), which completes the proof. \hfill
$\square$

A useful corollary to this Theorem for any specific $l$ can be
obtained by expanding $P_S^{(\ell)}$ in Eq.~\ref{theorem}.
In the case $\ell = 2$, it reads
\begin{eqnarray}
\overline{ \langle A \rangle_\psi \langle B \rangle_\psi} &=&
\sum_{ijmn}\frac{2 A_{ij} B_{mn} (P_S^{(2)})_{ji,nm}} {N^2+N}
\nonumber \\ &=& \sum_{ijmn}\frac{A_{ij} B_{mn}
(\delta_{ij}\delta_{mn} + \delta_{in}\delta_{mj})} {N^2+N}
\label{trace} \\ &=& \frac{\Tr\{A\}\Tr\{B\} +
\Tr\{AB\}}{N^2+N}.\nonumber
\end{eqnarray}
Similar expressions can be derived  for $\ell >2$, which
involves the properly normalized sum of all combinations of traces of
products and products of traces.

\begin{figure}[!tbh]
\centering
\includegraphics[width=8cm]{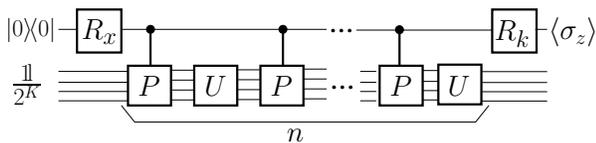}
\caption{Quantum circuit evaluating the average fidelity
  $\overline{F_n(\psi)}$ between the perturbed and unperturbed maps
  $U$ and $U_p = UP$.  The gates $R_k$ are $\pi/2$ rotation in the
  Bloch sphere around axis $k=x$ or $y$. When $k$ is set to $x$, we
  get the real part of $\Tr\{(U^n)^\dagger U_p^n \}/N$ while $k=y$
  yields the imaginary part. The unitary operator $P$ is applied
  conditionally: when the probe qubit is in state $\ket 1$, the
  unitary $P$ is applied to the lower register while no transformation
  is performed when the state of the probe qubit is $\ket 0$.}
\label{circuit}
\end{figure}

To arrive at our algorithm, it is sufficient to write the average
fidelity as 
$\overline{F_n(\psi)} = \overline{ \langle (U^n)^\dagger U_p^n \rangle_\psi \langle (U_p^n)^\dagger U^n \rangle_\psi}$,
and apply the identity from Eq.~\ref{trace} to obtain
\begin{equation}
\overline{F_n(\psi)} =
\frac{\left|\Tr\{(U^n)^\dagger U_p^n \}\right|^2 +N}{N^2+N}.
\label{Afidelity}
\end{equation}
The specific form of our theorem with $\ell =2$, unitary
$A$, and $B=A^\dagger$ was discovered by M., P., and
R. Horodecki~\cite{3H1999}, but our proof simplifies the presentation.
An efficient DQC1 algorithm to evaluate the trace of any unitary operator
[here, $(U^n)^\dagger U_p^n$], provided that it admits an efficient gate
decomposition, was presented in Ref.~\cite{MPSKLN2002}. If the perturbed 
map takes the form $U_p = UP$ for some unitary operator $P$ (e.g., $P
= \exp\{-i\delta V\}$ as above), the circuit can be further simplified
into the one illustrated on Fig.~\ref{circuit}.

We now analyze the complexity of our algorithm.  We assume that $U$
and $U_p$ admit $\epsilon$-accurate gate decompositions whose sizes
grow as $L(K,\epsilon) \in$ poly($K,1/\epsilon$). This implies that
the controlled version of these gates also scale as
$L(K,\epsilon)$~\cite{BBCDMSSSW1995}. We see from
Eq.~\ref{Afidelity} that the variance of $\overline{F_n(\psi)}$ is at
most twice the variance of $\Tr\{(U^n)^\dagger U_p^n \}/N$. Therefore,
the overall algorithm -- estimating $\overline{F_n(\psi)}$, to within
$\epsilon$, with error probability at most $p$ -- requires resources
growing as $L(K,\epsilon)n\log(1/p)/\epsilon^2$, so it is
efficient. (The range in $n$ over which the decay is studied should be
independent of the system's size.)  This algorithm thus provides an
{\em exponential} speed-up over all known classical procedures and
uses a {\em single} bit of quantum information. Furthermore, it
eliminates any cost of averaging the fidelity over a random set of
initial states, as this averaging is done directly.

In order to implement certain unitary maps on $K$ qubits efficiently,
it is necessary to introduce a number $K_a$ of ancillary qubits (a
``quantum work-pad'') in the fiducial state $\ket{\psi_0}$. Ancillary
qubits in pseudo-pure states can be used in the DQC1 setting.  As a
first step of the computation, part of the polarization of the probe
qubit of Eq.~\ref{eq:initial} can be transferred to ancillas initially
in maximally mixed states.  Thus, as long as the size $K_a$ of the
work-pad is at most poly-logarithmic in $K$, the algorithm remains
efficient.

Perhaps the most surprising feature of the quantum algorithm as it is
presented in Fig.~\ref{circuit} is that the probe never gets entangled
with the system throughout the computation. To show this, consider a
generalized version of the circuit of Fig.~\ref{circuit} where the
$P$'s and the $U$'s are free to differ at each iteration, i.e.~at step
$j$, we apply $P_j$ conditionally on the probe qubit, followed by
$U_j$. This generalization is necessary since the controlled $P$ gate
will in general be decomposed as a sequence of elementary controlled
and regular gates~\cite{BBCDMSSSW1995}.  Initially, the probe qubit is
in state $\alpha \ket 0 + \beta \ket 1$.  After $k$ steps, the state
of the QIP is
\begin{eqnarray}
\rho_k &=& \frac 1N \left\{ 
|\alpha|^2 \kb{0}{0}\otimes \II + \alpha\beta^* \kb{0}{1}
\otimes S^\dagger \right. \nonumber \\
&+& \left. \alpha^*\beta \kb{1}{0} \otimes S
+ |\beta|^2 \kb{1}{1}\otimes \II \right\}
\label{state_n}
\end{eqnarray}
where $S = U_kP_k\ldots U_2P_2U_1P_1 U_1^\dagger U_2^\dagger \ldots
U_k^\dagger$. Decomposing this state in the eigenbasis of the unitary
matrix $S\ket{\phi_j} = e^{is_j} \ket{\phi_j}$, we get
\begin{equation}
\rho_k = \frac 1N \sum_j \kb{\alpha_j}{\alpha_j} \otimes \kb{\phi_j}{\phi_j}
\end{equation}
where $\ket{\alpha_j} = \alpha\ket 0 + \beta e^{is_j} \ket 1$; the
state is separable. Its separability supports the point of view that the
power of quantum computing derives not from the special features of 
{\it quantum states} --- such as entanglement --- but rather from
fundamentally {\it quantum operations}~\cite{evolution}.
  
Our algorithm also illustrates why chaotic environments are expected
to produce decoherence more rapidly than integrable ones
\cite{bz2002}. Consider the probe qubit of Fig.~\ref{circuit} as a
quantum system interacting with a complex environment consisting of
$K$ two-level systems. After a ``time'' $n$, the state of the system
is given by tracing out the $K$ environmental qubits from
Eq.~\ref{state_n}. The diagonal elements of the reduced density matrix
$|\alpha|^2$ and $|\beta|^2$ are left intact while the off-diagonal
elements $\alpha\beta^*$ and $\alpha^*\beta$ are decreased by a factor
$|\Tr\{S\}|$ which is roughly equal to $\sqrt{\overline{F_n(\psi)}}$.
Thus, for an environment with chaotic dynamics, the system will
decohere at an exponential rate, whereas the rate of decoherence 
should be slower for non-chaotic environments. This analogy also
provides a very simple example of decoherence without
entanglement~\cite{EP2002}.

On the circuit of Fig.~\ref{circuit}, only the perturbation gates $P$
are conditioned on the state of the probe qubit. This suggests a dual
interpretation of the algorithm as quantum circuit and quantum probe.
On the one hand, $U$ could be a known unitary transformation which is
being simulated on the lower $K$-qubit register over which we have
universal control. Then, the gate $U$ would simply be decomposed as a
sequence of elementary gates as prescribed in Refs.~\cite{simul} for
example.  On the other hand, the lower register could be a real
quantum system undergoing its natural evolution $U$ which might not
even be known.  Then, the probe qubit should really be regarded as a
probe which is initialized in a quantum superposition, used to
conditionally {\it kick} the system, and finally measured to extract
information about the system under study. In this case, it is not
necessary to have universal control over the lower register (the
quantum system), we must simply be able to apply a conditional small
unitary transformation to it.

\begin{figure}[!tbh]
\centering
\includegraphics[width=5.5cm]{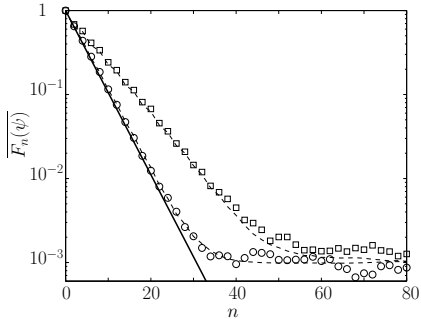}
\caption{Fidelity decay $F_n(\psi)$ averaged over 50 initial 
  computational basis states for $U_{QKT}$ in a regular regime ($k=1$, squares)
  and chaotic regime ($k=12$, circles). The dashed lines represent the
  exact average Eq.~\ref{Afidelity} and the full line shows the exponential
  decay at the Fermi golden rule rate $\Gamma$. }
\label{plot}
\end{figure}

Finally, Eq.~\ref{Afidelity} provides a useful numerical tool that can be used
to compute the {\em exact} average fidelity instead of estimating it by
averaging over a finite random sample of initial states. In
Ref.~\cite{EWLC2002}, fidelity decay was illustrated on the quantum
kicked top map $U_{QKT} = \exp\{-i\pi J_y/2\}\exp\{-ikJ_z^2/j\}$
acting on the $N=2j+1$ dimensional Hilbert space of angular momentum
operator $\vec{J}$. The chosen perturbation operator was $P =
\prod_{j=1}^K \exp\{-i\delta\sigma_z^j/2\}$, a collective
rotation of all $K$ qubits of the QIP by an angle $\delta$. The decay
rate (governed by the Fermi golden rule in this regime) was evaluated
to $\Gamma = 2.50\delta^2$ for this perturbation~\cite{EWLC2002}.
$\overline{F_n(\psi)}$ was estimated in both chaotic ($k=12$) and
regular ($k=1$) regimes of the kicked top by averaging over 50 initial
states. We reproduce these results on Fig.~\ref{plot} and compare them
with the exact average Eq.~\ref{Afidelity} and theoretical prediction
$e^{-\Gamma n}$. The random sample is in good agreement with the exact
average except that the former shows fluctuations.  Furthermore, the
decay in the chaotic regime is in excellent agreement with the Fermi
golden rule. While the decay is slower in the regular regime, it is
not clear from these results that it is not exponential.

We have presented an efficient quantum algorithm which computes the
average fidelity decay of a quantum map under perturbation using a
single bit of quantum information. The quantum circuit for this
algorithm establishes a link between decoherence by a chaotic
environment and fidelity decay. Using a special case of our theorem,
we have numerically evaluated the exact average fidelity decay for the
quantum kicked top, and found good agreement with previous
estimations using random samples. Although we have mainly motivated
our algorithm for the study of quantum chaos, we believe that it has
many other applications such as characterizing noisy quantum channels
and computing correlation functions for many-body systems. We have
also shown that our algorithm can be viewed as a special experiment
where a quantum probe is initialized in a superposition and used to
conditionally kick the system under study.  This type of {\it quantum
information science byproduct} might open the horizon to new types of
experimental measurements where a small QIP is used to extract
information from the quantum system under study.  Finally, the
effective speed-up despite the limited presence of entanglement --- in
particular its complete absence between the quantum probe and the
mixed register --- is a step forward in our understanding of the
origin of quantum-computational speed-up.

We thank J. Emerson, G. Milburn, J.P. Paz, and W.H.
Zurek for helpful discussions. 
We also acknowledge the Benasque Center for
Science where this work was initiated.
This work was supported in part by NSERC, ARDA, CIAR, ACI
s\'ecurit\'e informatique, and by the 
Department of Energy, under contract W-7405-ENG-36.


\begin{thebibliography}{99}

\bibitem{Peres1984} A. Peres,
{\it Phys. Rev. A} {\bf 30} 1610 (1984).

\bibitem{FD} R.A. Jalabert and H.M. Patawski, 
{\it Phys. Rev. Lett} {\bf 86} 2490 (2001); 
F. Cucchietti, C.H. Lewenkopf, E.R. Mucciolo, H. Patawski, and R.O. Vallejos,
arXiv: nlin.CD/0111051 (2001);
G. Benenti and G. Casati, arXiv: quant-ph/0112060 (2001);
T. Prosen and M. Znidaric,
{\it J. Phys. A} {\bf 35} 1455 (2002);

\bibitem{JSB2001} P. Jacquod, P.G. Silvestrov, and C.W.J. Beenakker,
{\it Phys. Rev. E} {\bf 64} 55203 (2001).

\bibitem{EWLC2002} J. Emerson, Y. Weinstein, S. Lloyd, and D. Cory,
{\it Phys. Rev. Lett.} {\bf 89} 284102 (2002).

\bibitem{BCD2002} R. Blume-Kohout, C.M. Caves, and I.H. Deutsch, 
{\it Foundations of Physics} {\bf 32}, 1641 (2002)

\bibitem{KL1998} E. Knill and R. Laflamme,
{\it Phys. Rev. Lett} {\bf 81} 5672 (1998).

\bibitem{Cory2000} D.G. Cory {et al.},
{\it Fortschr. Phys.} {\bf 48}, 875 (2000).

\bibitem{Huber1981} P.~J.~Huber, 
{\it Robust statistics} (Whiley, New York, 1981).

\bibitem{simul} R. Schack,
{\it Phys. Rev.} {\bf A 57}, 1634 (1998);
B. Georgeot and D.L. Shepelyansky,
{\it Phys. Rev. Lett.} {\bf 86}, 2890 (2001);
G. Benenti, G. Casati, S. Montangero, and
  D.L. Shepelyansky,
{\it Phys. Rev. Lett} {\bf 87}, 227901 (2001).

\bibitem{ELPC2003} J. Emerson, S. Lloyd, D. Poulin, and D. Cory,
arXiv: quant-ph/0308164 (2003).

\bibitem{PLMP2003} D. Poulin, R. Laflamme, G.J. Milburn, and J.P. Paz,
{\it Phys. Rev. A} {\bf 68} 022302 (2003).

\bibitem{Haake2001} F. Haake, 
{\it Quantum Signatures of Chaos}, Spriger-Verlag, Berlin (2001). 

\bibitem{ZHSL1998_BCJLPS1999} K. Zyczkowski, P. Horodecki, A. Sanpera, and M. Lewenstein,
{\it Phys. Rev. A} {\bf 58} 883 (1998); 
S.L. Braunstein, C.M. Caves, R. Jozsa, N. Linden, S. Popescu, and R. Schack,
{\it Phys. Rev. Lett.} {\bf 83} 1054 (1999).

\bibitem{EJ1998} A. Ekert and R. Jozsa,
{\it Phil. Trans. R. Soc. Lond. A} 1769 (1998);
Jozsa, R. in {\it The Geometric Universe} 
edited by S. Huggett, L. Mason, K. P. Tod, S. T. Tsou and N. Woodhouse, 
Oxford University Press (1998).

\bibitem{BBDEJM1997} A. Barenco, A. Berthiaume, D. Deutsch, A. Ekert, R. Jozsa, and C. Macchiavello,
{\it SIAM J. Comput.} {\bf 26} 1541 (1997).



\bibitem{3H1999} M. Horodecki, P. Horodecki, and R. Horodecki,
{\it Phys. Rev. A} {\bf 60} 1888 (1999).

\bibitem{MPSKLN2002} C. Miquel, J.P. Paz, M. Saraceno, E. Knill,
  R. Laflamme, and C. Negrevergne,
{\it Nature} {\bf 418}, 59 (2002).


\bibitem{BBCDMSSSW1995} A. Barenco, C.H. Bennett, R. Cleve, D.P.
 DiVincenzo, N. Margolus, P.W. Shor, T. Sleator, J.A. Smolin, and H.
 Weinfurter, {\it Phys. Rev.} {\bf A 52}, 3457 (1995).

\bibitem{evolution}  R. Laflamme, D.G. Cory, C. Negrevergne, and L. Viola,
{\it Quant. Info. and Comp.} {\bf 2} (2001);
D. Poulin,
{\it Phys. Rev. A} {\bf 65} 42319 (2002).

\bibitem{bz2002} R. Blume-Kohout and W. H. Zurek,
arXiv: quant-ph/0212153 (2002).

\bibitem{EP2002} J. Eisert and M.B. Plenio,
{\it Phys. Rev. Lett.} {\bf 89} 137902 (2002).

\end{thebibliography}
\end{document}